\documentclass
[aps,reprint,showkeys,showpacs,
pre,
preprintnumbers,superscriptaddress,amsmath,amssymb,amsfonts]{revtex4-1}

\usepackage{hyperref}
\usepackage[dvips]{graphicx}	% Include figure files
\usepackage{latexsym}
\usepackage{dcolumn}		% Align table columns on decimal point
\usepackage{bm}			% bold math
\usepackage{color}

\begin{document}

% ---------------------------------------------------------------------
% title page
% ---------------------------------------------------------------------
\title{%
Robustness of onion-like correlated networks
against targeted attacks
}

\author{Toshihiro Tanizawa}
\email[E-mail:]{tanizawa@ee.kochi-ct.ac.jp}
\affiliation{Kochi National College of Technology,
200-1 Monobe-Otsu, Nankoku, Kochi 783-8508, Japan}
\author{Shlomo Havlin}
\affiliation{Minerva Center and Department of Physics,
Bar-Ilan University, 52900 Ramat-Gan, Israel}
\author{H.~Eugene Stanley}
\affiliation{Center for Polymer Studies and Department of Physics,
Boston University, Boston, MA 02215, USA}

\date{\today}

% --------------------------------------------------------------------
% abstract
% --------------------------------------------------------------------
\begin{abstract}
 Recently, it was found by Schneider et al.\ [Proc.\ Natl.\ Acad.\ Sci.\
 USA, {\bf 108}, 3838 (2011)], using simulations, that scale-free
 networks with ``onion structure'' are very robust against targeted high
 degree attacks.  The onion structure is a network where nodes with
 almost the same degree are connected.  Motivated by this work, we
 propose and analyze, based on analytical considerations, an onion-like
 candidate for a nearly optimal structure against simultaneous random
 and targeted high degree node attacks.  The nearly optimal structure
 can be viewed as a set of hierarchically interconnected random regular graphs,
 the degrees and populations of whose nodes are specified by the degree
 distribution.  This network structure exhibits an extremely assortative
 degree-degree correlation and has a close relationship to the ``onion
 structure.''  After deriving a set of exact expressions that enable us
 to calculate the critical percolation threshold and the giant component
 of a correlated network for an arbitrary type of node removal, we apply
 the theory to the cases of random scale-free networks that are highly
 vulnerable against targeted high degree node removal.  Our results show
 that this vulnerability can be significantly reduced by implementing
 this onion-like type of degree-degree correlation without much
 undermining the almost complete robustness against random node removal.
 We also investigate in detail the robustness enhancement due to
 assortative degree-degree correlation by introducing a joint
 degree-degree probability matrix that interpolates between an
 uncorrelated network structure and the onion-like structure proposed
 here by tuning a single control parameter.  The optimal values of the
 control parameter that maximize the robustness against simultaneous
 random and targeted attacks are also determined.  Our analytical
 calculations are supported by numerical simulations.
\end{abstract}

% -------------------------------------------------------------------
% PACS (the Physics and Astronomy Classification Scheme)
% -------------------------------------------------------------------
				    

\maketitle

% --------- Title page ends -----------------------------------------

% ======================================================================
%  main body
% ======================================================================

\section{\label{sec:intro}Introduction}

Many complex systems in real world can be modeled by complex
networks \cite{Watts:1998vz,Barabasi:1999uu,Albert:2002wu,Newman:2003wd,%
Dorogovtsev:2003Evolut_Networ,Cohen:2005wq,%
Boccaletti:2006gb,Calderelli:2007Large_Scale_Struct,
newman:2010_networ,cohen:2010_compl_networ,Gao:2011fq%
}.
Generally speaking, the cooperative performance of complex systems
fundamentally relies on the global connectivity of their components.
These complicated systems are, however,
usually placed in an ever-changing external
environment where the components or the connections could
be constantly added, eliminated, or changed.
Such changes may potentially affect the global connectivity
of the network under consideration
to the extent in which the global connectivity
could be completely lost and the system represented
by the network will lose its functionality.  The analysis of the response of
the global connectivity caused by the alteration of the network,
or targeted attacks, has been therefore
one of the main issues of the complex network analysis.

Most of the existing theoretical studies on the robustness of complex
networks depend only on the degree distribution
\cite{%
Molloy:1995tw,%
albert:2000_error,Cohen:2000vq,Moore:2000ve,%
Callaway:2000vd,Cohen:2001hf,Cohen:2002br,Chung:2002wu,Schwartz:2002dn,Gallos:2005gs,%
Shargel:2003cu,Vazquez:2003fq,
Paul:2004br,Tanizawa:2005fd,Tanizawa:2006dn,%
Paul:2006cd,Donetti:2006ck,Paul:2007br,buldyrev:2010_catas,Huang:2011cg,%
Dorogovtsev:2003Evolut_Networ,Cohen:2005wq,%
Boccaletti:2006gb,Calderelli:2007Large_Scale_Struct,
newman:2010_networ,cohen:2010_compl_networ,Gao:2011fq%
}.
However, as noted by Newman,
networks in real world exhibit rather strong tendency, or correlation,
in the connection between nodes of different degrees \cite{Newman:2002jj}.
He introduced the terms, assortative and disassortative correlations,
to describe the tendency of nodes in a network
to make connections between the same degree and
between different degrees, respectively,
and calculated how the giant component collapse
for specific kinds of correlated networks
against random node removal.
Newman applied
the generating function formalism and showed
the enhancement of the resiliency of
networks with assortative degree-degree correlation.

In addition to Newman's pioneering work,
there are few theoretical works on robustness analysis
including degree-degree correlations~%
\cite{Serrano:2006ik,Goltsev:2008bf,Shiraki:2010is,Ostilli:2011hb}.
Among these, Goltsev et al.\ focused on the
evaluation of critical exponents of correlated complex networks
in the vicinity of node percolation transition for the case of
random node removal \cite{Goltsev:2008bf}.
Here we extend their formalism and proceed
to the robustness analysis of a correlated complex network
against arbitrary types of node removal.

Recently, Schneider et al.\ developed an interesting numerical approach
for enhancing network robustness against high degree node removal %
\cite{Schneider:2011ip,Herrmann:2011hd}.
They start from an uncorrelated random network
with a given degree distribution.
Next, they randomly choose two pairs of links
and exchange the destinations of the two links
between them keeping the overall degree distribution unchanged.
If this exchange improves the robustness of the network
against targeted node removal, the exchange is accepted.
By repeating this procedure,
the robustness of the network is enhanced step by step.
They applied this method to several types of networks
with broad degree distributions
and found that the final robust networks have
a common ``onion-like'' topology consisting
of a core of highly connected nodes hierarchically
surrounded by rings of nodes with
decreasing degree \cite{Schneider:2011ip,Herrmann:2011hd}.
In each ring most of the nodes are of the same degree.
A numerical method that improves the convergence
to the onion structure is reported by Wu and Holme recently \cite{Wu:2011uw}.

Motivated by the onion-like topology, we study here analytically
the robustness of a family of such systems.
In our approach we obtain analytical expressions for the critical threshold
and for the giant components,
where the degree-degree correlation is fully incorporated.
Due to the analytical approach,
a statistical treatment over large number of realizations
as done in computer simulations is not needed
to obtain definite results.
Nevertheless both analytical and simulation approaches are
necessary and complementary, in particular,
for testing the analytical approach.
Interestingly, the optimal structure we find here against simultaneous random
and targeted high degree node removals is very similar
to the ``onion-like'' structure
found by Schneider et al.\ \cite{Schneider:2011ip}.
The optimal structure obtained consists of hierarchically
and weakly interconnected random regular graphs.

The paper is organized as follows.
In Section \ref{sec:theory}，
we derive a set of analytical equations for
the critical node threshold
and the giant component fraction for an arbitrary type
of node removal where the degree-degree correlation is fully incorporated.
We begin our analysis in Section \ref{sec:heuristic}
by presenting a simple theoretical argument
to derive the optimal network structure
against targeted high degree node removal.
In Section \ref{sec:analysis sep reg graphs},
we describe the properties of a set of separated random regular graphs
as the first step to understand the structure of
the optimal network described in the previous section.
Support of our analytical results by numerical simulations is also presented.
In Section \ref{sec:interconnected}，
we analyze the properties of interconnected random regular graphs
of different degrees $k$
by introducing a joint degree probability matrix
that can describe the transition between
a set of separated random regular graphs
and a completely uncorrelated single random network
by tuning a single control parameter under the condition of
having a fixed degree distribution.
The optimal values of the control parameter
that maximize the robustness against
simultaneous random and targeted high degree node attacks is determined.
In Section \ref{sec:summary} we summarize the results.

\section{\label{sec:theory}Theory}

We start from the joint degree-degree probability matrix, $P(k, k')$,
which is the probability that a randomly chosen link emanates
from a $k$-degree node and ends at a $k'$-degree node.
In this article, we consider only the cases of undirected networks,
where the symmetry $ P(k, k') = P(k', k) $ holds.
The sum of $ P(k, k') $ over $k'$ is the probability
that a randomly chosen link starts from a $k$-degree node.
It is related to the probability density of the degree distribution,
$P(k)$, through the relation,
$ \sum_{k'} P(k, k') = k P(k)/ \langle k \rangle$,
where $\langle k \rangle$ is the average degree.
By definition, $ \sum_k P(k) = 1$. Note that
the sum $\sum_{k'} P(k, k')$ has to be fixed
if we fix the degree distribution, $P(k)$.
The conditional probability, $ P(k'|k) $,
that a randomly chosen link emanating from
a $k$-degree node leads to a $k'$-degree node is defined by
$ P(k'|k) \equiv P(k, k')/\sum_{k'} P(k, k') = P(k, k')/\left( k P(k)/ \langle k \rangle \right)$.

When the nodes of a network are removed according to the degree of nodes,
the remaining fraction of $k$-degree nodes is reduced
by a factor $b_k \: (0 \le b_k \le 1)$
from the original fraction, $P(k)$.
The total remaining fraction of nodes, $p$,
is calculated as $p = \sum_k b_k P(k) $.

The giant component in a complex network is a cluster of connected
nodes, where its normalized size in the network, $S$,
remains finite as the total number of nodes, $N$, becomes infinite.
Non-zero values of $S$ indicate a macroscopic connectivity
of the network under consideration.

To calculate the critical value of the remaining fraction of nodes, $p_c$,
above which the giant component, $S$, begins to take a non-zero value,
we extend the generating function method %
\cite{Callaway:2000vd, Goltsev:2008bf}
by incorporating the degree-degree correlation
under an arbitrary way of node removal.
Let $x_k$ be the probability that a randomly chosen link
from a $k$-degree node does not lead to the giant component.
Under the condition that the network only consists of trees,
which is justified in the limit of $N \to \infty$,
the probabilities, $x_k,\; \left( k = m, m+1, \dots, K \right)$,
for non-zero values of $b_k$,
and the node fraction of the giant component, $S$, are determined by
the following set of equations:
\begin{align}
 x_k &= 1 - \sum_{k'} b_{k'} P(k'|k) + \sum_{k'} b_{k'} P(k'|k) \left( x_{k'} \right)^{k'-1} \label{eq: xk}\\
 S &= p - \sum_k b_k P(k) \left( x_k \right)^k = \sum_k b_k P(k) \left( 1 - (x_k)^k\right).
 \label{eq: S}
\end{align}
Obviously, $x_k = 1$ for removing all $k$-degree nodes ($b_k = 0$).
Note that these equations contain
the remaining fraction of $k$-degree nodes, $b_k$.
Equations (\ref{eq: xk}) and (\ref{eq: S}) are a necessary extension
of existing works in order to investigate all types of node removal.
The degree-degree correlation is included in the conditional probability,
$P(k'|k)$.

Below the critical remaining fraction of nodes, all $x_k$'s are equal to one
and it follows from Eq.~(\ref{eq: S}) that $S = 0$ (no giant component).
At criticality where the giant component emerges, at least one of $x_k$'s
takes a value slightly smaller than one.
In the vicinity of the critical point, we assume $x_k = 1 - y_k$ and
expand Eq.~(\ref{eq: xk}) in terms of infinitesimally positive quantities
$y_k$. The equation obtained by this expansion becomes
\begin{equation}
 y_k = \sum_{k'} B_{kk'} y_{k'} + O(y_k^2),\label{eq:branching_mat}
\end{equation}
where the ``branching matrix,'' $B_{kk'}$, is defined by
$B_{kk'} \equiv b_{k'} P(k'|k) (k'-1)$.
The eigenvalues of the branching matrix are all non-negative and
can be ordered according to their values.
The critical point can be obtained by the point at which
the largest eigenvalue of $B_{kk'}$ becomes unity \cite{Goltsev:2008bf}.

As we will show in Section \ref{sec:analysis sep reg graphs},
we can also calculate the critical exponent
$\beta$ defined by $S \sim \left\vert p - p_c \right\vert^\beta$
in the vicinity of the critical point.

\section{the Optimal Structure}
\label{sec:heuristic}

Let us begin by reviewing the robustness of random regular graphs,
which are networks that consists of only nodes with the same degree, $k$.
Such networks serve as components in the structure studied in this paper.

Since all nodes in a random regular graph have the same degree, $k$,
there is no difference between random and targeted attacks.
It is well known from percolation theory that a random regular graph with
degree $k \ge 2$ contains a single giant component when the remaining
node fraction after random removal of nodes
exceeds the critical threshold $p_c = 1/(k - 1)$.
It should be pointed out that the critical node threshold
for the random regular graph for $k = 2$ is one.
This means that the giant component for a random regular graph with $k = 2$
is always critical and collapses as soon as a single node is removed.
Because of this criticality, the giant component fraction, $ S $, of a $k = 2$
random regular graph is not unity but close to 0.8 \cite{Bollobas:1984ti}.

The robustness of a given network depends on the method of node removal.
For example,
scale-free networks are almost completely robust against
random node removal while they are extremely
vulnerable against targeted removal of high degree nodes
\cite{albert:2000_error,Cohen:2000vq,Callaway:2000vd,Cohen:2001hf}.
The results for the robustness is, however,
derived for random networks and thus
are based only on the degree distribution.
It is interesting, therefore, to clarify to what extent
we are able to improve the robustness of a complex network against
targeted attack by introducing the degree-degree correlation while
keeping the network degree distribution unchanged.

With this in mind,
we focus on the improvement of the robustness of complex networks
against targeted high degree node attack.
We limit our analysis to networks where the number of
$k$-degree nodes decreases with increasing $k$.
In targeted high degree attack, all nodes that have higher degrees than
a certain value are eliminated.
Removing a node also eliminates all the edges attached to it.
Since the edges are connected with the remaining lower degree nodes,
the elimination of those edges undermines
the global connectivity of the remaining
lower degree node component.
In order to minimize such undermining effects as much as possible,
the number of edges that connect removed higher degree nodes and the
remaining lower degree nodes should be minimized as much as possible.
Hence the following requirement should be fulfilled.

\vspace{10pt}

\noindent
{\bf{Requirement}:} The $k$-degree nodes should not be connected to
nodes with degree, $k'$, lower than $k$ $(k' < k)$.

\vspace{10pt}

This Requirement yields that
most of the edges should connect nodes of the same degree.
Thus the optimal structure built up from a set of random regular (RR) graphs
naturally emerges.
To form an entirely connected single network,
these RR graphs must be connected with one another.
The most robust network against targeted attack
with a given degree distribution can, therefore,
be constructed by the following procedure.

\begin{enumerate}
 \item
      Prepare a suitable number of nodes for each degree according to
      the given degree distribution.
      We assume that the number of nodes for each degree is
      so large that all edges can find nodes to be attached in
      both end points.

 \item
      Let the smallest degree be $m$ and
      begin to construct the network from an $m$-degree component,
      which is the last remaining component
      for targeted high degree node removal.
      If the Requirement is completely fulfilled,
      no edges of the $m$-degree component are eliminated by targeted removal
      of nodes with degree larger than but not equal to $m$.
      The last remaining $m$-degree component forms, therefore,
      an RR graph of degree $m$.

 \item
      Next, attach the nodes with degree $ m+1 $.
      According to the Requirement,
      the attached $(m+1)$-degree nodes cannot be connected
      to the (smaller) $m$-degree component.
      Thus all $(m+1)$-degree nodes should be connected
      with one-another and forms
      an RR graph of degree $(m + 1)$.
      
      Up to this point, the network consists of two separated RR graphs with
      degree $m$ and $m+1$. However, to make a single connected network
      we have to connect these two components.
      To fulfill the Requirement as much as possible under the condition of
      the fixed degree distribution, we break two edges, the one of which is
      in the RR graph of degree $m$ and the other of
      which is in the RR graph of
      degree $m+1$, and rewire these two edges.
      Note that this rewiring does not change the degree distribution.
      
 \item
      Attaching the nodes with next larger degree, $m + 2$ can be performed
      in the same way.
      First, following  the Requirement,
      these nodes should be connected with one-another.
      Hence, an RR graph with degree $m + 2$ emerges.
      Next, to make a single connected network under the conditions of
      the Requirement and the fixed degree distribution,
      two edges in the RR graph of degree $m+1$
      and the RR graph of degree $m+2$ are broken and rewired.
      
      By repeating this argument up to the nodes
      with the largest degree, $K$,
      we reach the structure in which RR graphs with degrees
      hierarchically up from $m$ to $K$ are minimally interconnected.
      This structure has a close resemblance
      with the robust ``onion-like'' structure found using numerical simulations
      by Schneider et al.\ \cite{Schneider:2011ip,Herrmann:2011hd}.
\end{enumerate}

In the following Sections, we investigate the properties of this structure,
which we also refer to as the ``onion-like'' structure.

\section{Analysis of Separated Random Regular Graphs}
\label{sec:analysis sep reg graphs}

\begin{figure}
 \includegraphics[clip,width=4cm]{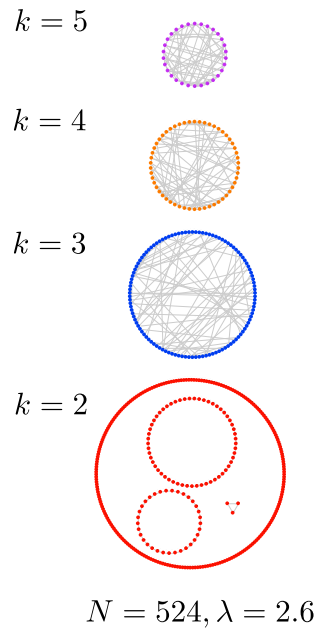}
 \caption{(Color online)
 An example of a set of
 separated random regular graphs from $k = 2$ to $k = 5$.
 The total node number $N = 524$ and the node number
 for each random regular graph is
 determined by the power-law degree distribution,
 $P(k) \propto k^{-\lambda}$ with $\lambda = 2.6$.
 Notice that the graph for $k = 2$ is composed of separated rings,
which can be fragmented by removing of a tiny amount (zero fraction) of nodes.
 This means that the giant component of the random regular graph for $k = 2$
 is always at the edge of criticality, as indeed predicted by Eq.~(\ref{eq_RR_threshold}).}
 \label{fig:sep reg}
\end{figure}

Let us begin with the case of a correlated network specified
by a delta function-like joint degree probability matrix,
\begin{equation}
 P(k, k') = \frac{k P(k)}{\langle k \rangle} \delta_{k k'}.\label{eq: extrm assort}
\end{equation}
This joint degree probability matrix leads to the conditional probability
that is a complete delta function: $P(k'|k) = \delta_{k k'}$.
In this network, only the nodes with the same degree are connected.
The whole network is therefore a set of random regular (RR) graphs
of all degrees of nodes from $m$ to $K$
where each degree fraction is specified by
the degree distribution $P(k)$.
In Fig.\ \ref{fig:sep reg}, we show an example of a set of RR graphs
specified by Eq.\ (\ref{eq: extrm assort}).

Since the onion structure proposed in the previous Section consists of
minimally connected RR graphs,
we expect that the properties of this structure should be
almost identical to a set of separated RR graphs
described by Eq.\ (\ref{eq: extrm assort}).
This is one of the reasons we begin our analysis
with a set of separated RR graphs.

% %%%%%%%%%%%%%%%%%%%%%%%%%%%%%%%%%
%  Figure 1
% %%%%%%%%%%%%%%%%%%%%%%%%%%%%%%%%%
% %%%%%%%%%%%%%%%%%%%%%%%%%%%%%%%%%%

Strictly speaking, there is no global giant component
in this network structure, since all RR graphs are separated.
We assume that the sum of the giant components of each RR graph
to be the ``virtual'' giant component in this case.
This definition of the ``virtual'' giant component naturally reflects
the ``real'' giant component when we add a minimal number of connections, or ``bridges,''
between these RR graphs.

The branching matrix, $B_{kk'}$, for this set of RR graphs
is diagonal with the diagonal elements, $B_{kk} = b_k (k - 1)$,
which are identical to the eigenvalues of the branching matrix
(See Eq.~(\ref{eq:branching_mat})).
Thus each RR graph with degree $k$ contained in this network
becomes critical when the remaining fraction of this mode, $b_k$ takes
the critical value
\begin{equation}
 b^*_k = \frac{1}{k - 1} \quad ( k \ge 2 ).\label{eq_RR_threshold}
\end{equation}
When nodes are removed starting from the highest to lower degrees,
the $b_k$'s of the removed degrees
become zero. The last remaining fraction of nodes is
that of the minimum degree, $m$, and the disappearance of
the giant component for this minimum degree indicates the collapse
of the finally remaining ``virtual'' giant component.
This occurs at
\begin{equation}
 b_m = b^*_m = \frac{1}{m - 1}, \quad b_k = 0 \;(k > m).
\end{equation}
Thus the node removal threshold for targeted attack
on high degree nodes of this extremely
assortative network becomes
\begin{equation}
 p_c = b^*_m P(m) = \frac{P(m)}{m - 1}.
  \label{eq:pc}
\end{equation}

At the emergence of the giant component, when $p \gtrsim p_c$,
the node fraction of the giant component, $S$, is characterized by
the critical exponent,
$\beta$, as $S \sim \left\vert p - p_c \right\vert ^\beta$.
Using the exact equations, Eqs.\ (\ref{eq: xk}) and (\ref{eq: S}),
we can also evaluate the value of $\beta$ as follows.

We note again that the case, $m = 2$, is a little tricky.
In the first place,
as we can see from Eq.\ (\ref{eq:pc}), the giant component
of the regular graph of the smallest degree does not emerge
until all the nodes with degree two are filled ($b_2 = 1$).
In other words, the giant component suddenly disappear as soon as
any single node of the smallest degree two is removed.
Thus the percolation transition in this case is discontinuous
and a finite value of $S$ suddenly appear at $p_c = P(2)$.
Second, the giant component for $k=2$ is always critical
and only about 80\% of the nodes of this smallest degree participate
in the giant component \cite{Bollobas:1984ti}.
The rest of the nodes of $k=2$ form
tiny rings, which do not contribute to the giant component.

For $m \ge 3$, the transition is continuous at $p_c = P(m)/(m - 1)$.
In this case, we can also evaluate the critical exponent $\beta$ as follows.
For $p \gtrsim p_c$, non-trivial solutions for Eq.~(\ref{eq: xk}),
which is
\begin{equation}
 x_k = 1 - b_k \left( 1 - (x_k)^{k-1}\right)
  \label{eq:xk assort}
\end{equation}
in this case, emerge.
Let us assume $x_m = 1 - \varepsilon$, where $\varepsilon \gtrsim 0$,
in the equation for the lowest degree $m$, and
expand Eq.\ (\ref{eq:xk assort}) up to the second order of $\varepsilon$.
This gives
\begin{equation}
 \varepsilon = \frac{2}{m - 2}\left\{ 1 - \frac{1}{b_m \left(m - 1 \right)}\right\}
  + O\left( \varepsilon^2 \right).\label{eq:epsilon}
\end{equation}
Since only the sub-graph of the lowest degree $m$ exists at the criticality,
the remaining node fraction is $p = b_m P(m)$;
therefore $p - p_c = \left( b_m - b_m^*\right) P(m)$.
Notice that $b_m^* = 1/\left(m - 1\right)$.
Thus we can rewrite Eq.\ (\ref{eq:epsilon}) as
\begin{equation}
 \varepsilon = \frac{2}{b_m^* P(m)\left(m - 2\right)}\left\vert p - p_c \right\vert
  + O\left( \left\vert p - p_c \right\vert^2 \right).
\end{equation}
Together with Eq.\ (\ref{eq: S}), which is
\begin{equation}
 S = b_m P(m) \left(1 - (x_m)^m \right)
\end{equation}
in this extremely assortative case,
we obtain
\begin{equation}
 S = \frac{2m}{m - 2}\left\vert p - p_c \right\vert + O\left( \left\vert p - p_c \right\vert^2\right).
\end{equation}
This means $\beta = 1$.
It is interesting that in the limit $m \to \infty$,
\begin{equation}
 S \approx 2 \left\vert p - p_c \right\vert \;\; \left( m \to \infty \right)
\end{equation}
in the vicinity of the transition.

To verify the above theoretical arguments,
we compare the size of ``virtual'' giant component obtained from theory
with the ``real'' giant component of the onion structure
proposed in the previous Section obtained by numerical simulation.
We focus on the cases of networks with
a power-law degree distribution
(scale-free networks), where
the degree distribution is represented by
$P(k) \propto k^{-\lambda}$.

\begin{figure*}
 \includegraphics[clip,width=12cm]{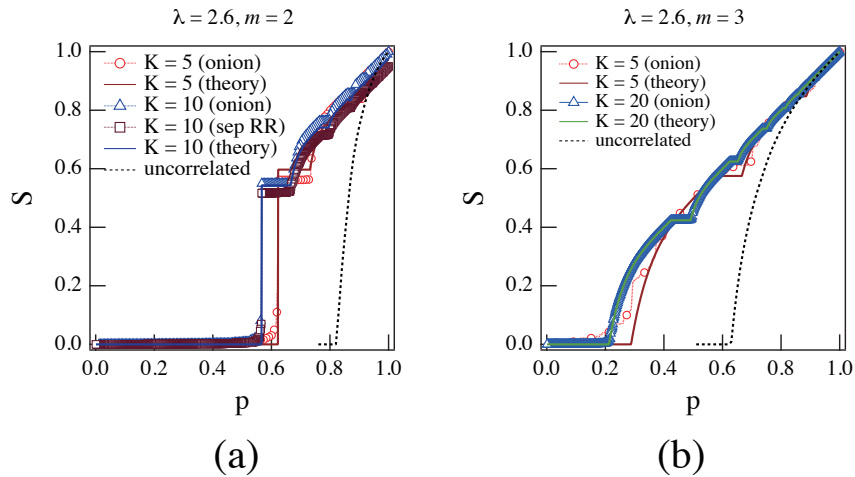}
 \caption{\label{fig:extreme_assort_targ_attack} (Color online)
 Plots of the values of the ``virtual'' giant component,
 $S$, as a function of the remaining fraction of nodes, $p$, obtained theoretically
 for extremely assortative scale-free networks specified
 by Eq.\ (\ref{eq: extrm assort})
 for various values of the maximum degree $K$.
 The degree distribution is fixed to the power-law $k^{-\lambda}$
 with an exponent $\lambda = 2.6$:
 (a) The giant component $S$ for $m = 2$ and (b) $S$ for $m = 3$.
 The values of $S$ for networks with the onion structure
 obtained by both theory and numerical simulations
 as well as the giant component fraction of
 the corresponding uncorrelated random scale-free network
 for $K = 5$ are shown.
 In (a), the values of $S$ for completely separated RR graphs
 (with no ``bridges'') obtained
 by simulation are also added for $K = 10$ for comparison.
 Notice that the difference between the values for the onion structure
 and those of the separated RR graphs are small for $m = 2$ and $K = 10$.
 For $m = 3$, this difference is indistinguishable.
 }
\end{figure*}

The results are shown in Fig.\ \ref{fig:extreme_assort_targ_attack}
for scale-free networks with $\lambda = 2.6$.
For the total number of nodes, $N$, in these simulations,
we take $N \approx 900$ for $K = 5$ and $N \approx 12000$ for $K = 10$
with $m = 2$ and $N \approx 900$ for $K = 5$
and $N \approx 20000$ for $K = 20$ with $m = 3$.
The agreement is excellent and thus supporting the analytical results
based on Eqs.~(\ref{eq: xk}) and (\ref{eq: S}).
Note, as seen in Fig.\ref{fig:extreme_assort_targ_attack},
that the value of the maximum degree $K$ does not play
an important role for the improvement of the robustness against targeted attack.
This can be understood since the highest degree nodes represent a low fraction of
the network nodes and are selectively removed in the targeted attack.%
\footnote{%
In order to verify the insensitivity of the robustness
to the maximum degree $K$, all the numerical calculations of the analytical expressions
in this article were also performed for $K = 500$ with the
same values for other parameters. The reason for taking small values of $K$
is to save time in the simulations for producing correlated networks
}

% %%%%%%%%%%%%%%%%%%%%%%%%%%%%%%%%%%%%%%
% Figure 2
% %%%%%%%%%%%%%%%%%%%%%%%%%%%%%%%%%%%%%%
% %%%%%%%%%%%%%%%%%%%%%%%%%%%%%%%%%%%%%%%

We also show in these figures the values of the giant component fraction
of the uncorrelated (random) network with the same degree distribution
% $P(k) \propto k^{-\lambda}$,
as a function of the remaining nodes.
It is clear that the strong assortativity,
obtained using the construction principle described in
Sec.~\ref{sec:heuristic}, considerably improves the
robustness of the scale-free network
against targeted high degree node removal.

\section{A Model for Interconnected Random Regular Graphs}
\label{sec:interconnected}

\subsection{The joint degree matrix}
\label{ssec:joint deg mat}

To investigate the effect of connections between random regular (RR) graphs
and to study the robustness of onion-like structures analytically,
we propose the following model
where the joint degree-degree probability matrix is defined by
\begin{equation}
 P(k, k') \propto \frac{k P(k)}{\langle k \rangle} \frac{k' P(k')}{\langle k \rangle}
  \exp \left[ - \frac{\left(k - k'\right)^2}{\sigma^2}\right],\label{eq: trans joint deg mat}
\end{equation}
which is normalized by the conditions
$\sum_{k'} P(k, k') = k P(k)/\langle k \rangle$ and $\sum_{k k'} P(k, k') = 1$.
This matrix, Eq.~(\ref{eq: trans joint deg mat}), contains a control parameter, $\sigma$.
In the limit, $\sigma \to 0$, the joint degree-degree matrix is that of
separated RR graphs, Eq.\ (\ref{eq: extrm assort}),
and in the limit, $\sigma \to \infty$,
it approaches to that of a completely random single uncorrelated network.
Note that for any value of $\sigma$, the degree distribution,
$P(k)$, is fixed and given
by the sum rule, $\sum_{k'} P(k, k') = k P(k)/\langle k \rangle$.

\begin{figure}
 \includegraphics[clip,width=7cm]{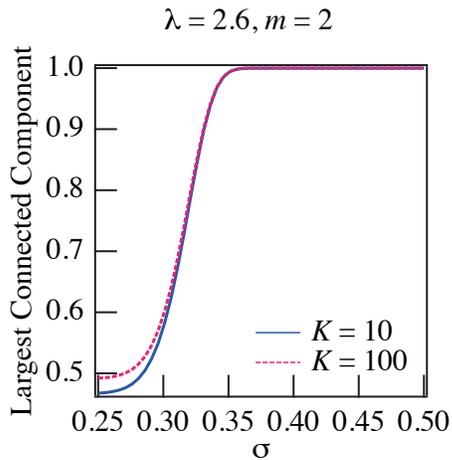}
 \caption{(Color online)
 The fraction of the largest connected component
 of scale-free networks with the exponent $\lambda = 2.6$,
 minimum degree $m = 2$, and maximum degrees $K = 10,\, 100$
 as a function of the parameter, $\sigma$, that controls the assortativity
 of the degree-degree correlation.
 Above $\sigma \approx 0.35$, the separated $k = 2$ components
 become connected to the larger degree components
 and the largest connected component spans the entire network.
 }
 \label{fig:init_GC}
\end{figure}

As mentioned earlier,
an RR graph for $k = 2$ generally consists of many separated rings.
Therefore, it is expected that until the control parameter for connection
between RR graphs, $\sigma$, obtains a suitably large value,
the largest connected component, which belongs to the $k  = 2$ component,
is not firmly connected to the larger degree ($k > m$) components.
In Fig.\ \ref{fig:init_GC},
we show the fraction of the largest connected component of a scale-free network
specified by Eq.\ (\ref{eq: trans joint deg mat})
with the exponent $\lambda = 2.6$,
minimum degree $m = 2$, and maximum degrees $K = 10 {\text{ and }} 100$
as a function of $\sigma$, which is obtained by numerical calculations
using the analytical expressions, Eqs.\ (\ref{eq: xk}) and (\ref{eq: S}).
From this Figure, we can see that for $m = 2$ we
need $\sigma \gtrsim 0.35$ for the largest connected component
to span the entire network.
For $m \ge 3$, we find (not shown) that
the largest connected component always spans the entire
network for any non-negative value of $\sigma$.

% %%%%%%%%%%%%%%%%%%%%%%%%%%%%%%%%%%%%%%%%%%%
% Figure 3
% %%%%%%%%%%%%%%%%%%%%%%%%%%%%%%%%%%%%%%%%%%%
% %%%%%%%%%%%%%%%%%%%%%%%%%%%%%%%%%%%%%%%%%%%%

\subsection{The critical percolation threshold}
\label{ssec:threshold}

The critical percolation threshold, $p_c$, is the minimum value
of the remaining node fraction required for a unique giant component
to be of the order of the entire network under a given way of node removal.
The threshold $ p_c $ is a useful measure of
the structural robustness of the network.
A smaller value of $p_c$ means that the network is more robust,
since we need to remove more nodes in order to destroy the giant component.

We calculate $p_c$ using the theoretical framework described
in Section \ref{sec:theory}.
For a given way of node removal, the critical point for
the vanishing giant component is specified by the point at which
the largest eigenvalue of the branching matrix
$B_{kk'} = b_{k'} P(k'|k)(k'-1)$
becomes unity. (See Eq.\ (\ref{eq:branching_mat}).)

\begin{figure*}
 \includegraphics[clip,width=13cm]{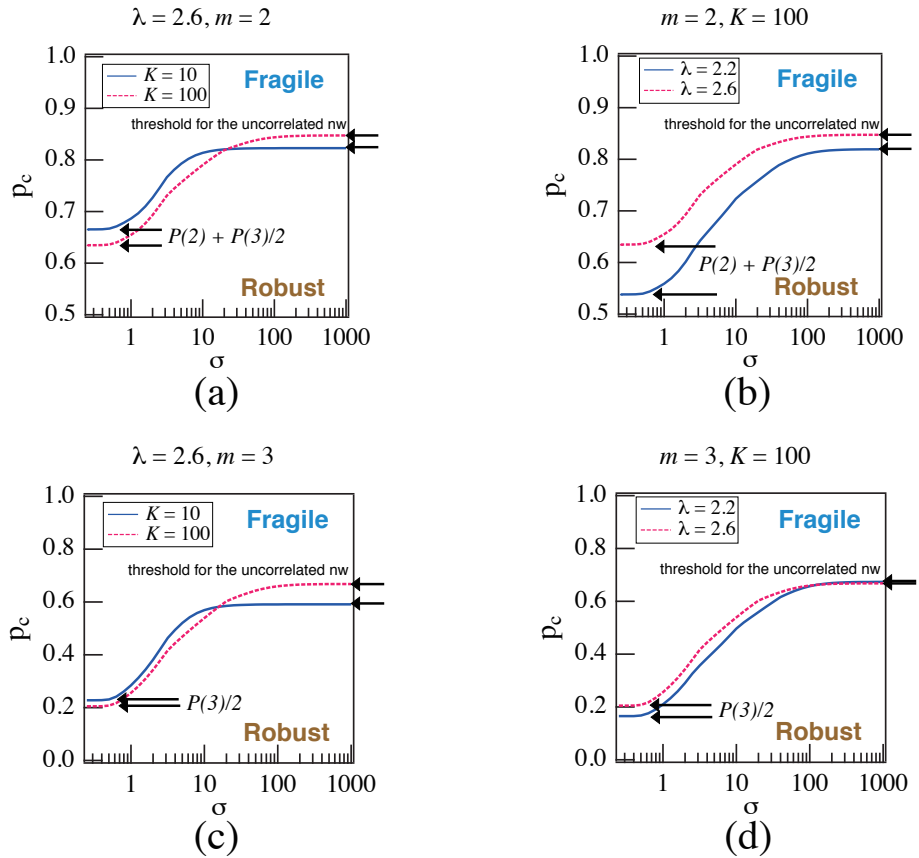}
 \caption{(Color online) Plots of the critical node threshold
 against targeted high degree node attack
 as a function of the parameter $\sigma$ that controls the assortativity of
 the degree-degree correlation (See Eq.~(\ref{eq: trans joint deg mat}).)
 for several scale-free networks.
 (a) Scale-free networks with the exponent $\lambda = 2.6$
 and minimum degree $m = 2$. The (blue) solid curve is for a network
 with maximum degree $K = 10$ and the (red) dotted curve is for $K = 100$.
 (b) Scale-free networks with $m = 2$ and $K = 100$.
 The (blue) solid curve is for $\lambda = 2.2$ and the (red) dotted curve
 is for $\lambda = 2.6$.
 (c) Scale-free networks with $\lambda = 2.6$ and $m = 3$.
 The (blue) solid curve is for $K = 10$ and the (red) dotted curve is for
 the network with $K = 100$.
 (d) Scale-free networks with $m = 3$ and $K = 100$.
 The (blue) solid curve is for $\lambda = 2.2$ and the (red) dotted curve
 is for $\lambda = 2.6$.
 For small values of $\sigma$, where the network structure approaches to
 be composed of weakly interconnected RR graphs, all critical node thresholds
 approach to $P(2) + P(3)/2$ for $m = 2$ and to $P(3)$ for $m = 3$.
 For all these cases, the giant component spans
 the entire network at the beginning of the node removal.
 Note that a smaller value of $ p_c $ generally means
 a more robust network structure.
 }
 \label{fig:eigen_values}
\end{figure*}

In Fig.\ \ref{fig:eigen_values}, we plot the threshold, $p_c$,
as a function of $\sigma$ for several scale-free networks.
For all calculations, the values of $p_c$ deviate from
those of the uncorrelated networks (seen at larger $\sigma$)
and become smaller (more robust)
as $\sigma$ decreases and the strong assortative
degree-degree correlation sets in.
Finally the values of $p_c$ converge to $P(2) + P(3)/2$ for
the networks with $m = 2$ and to $P(3)/2$ with $m = 3$
in the range $\sigma \lesssim 1$, where $m$ is the minimum degree.

% %%%%%%%%%%%%%%%%%%%%%%%%%%%%%%%%%%%%%%%%%%%%%
% Figure 4
% %%%%%%%%%%%%%%%%%%%%%%%%%%%%%%%%%%%%%%%%%%%%%
% %%%%%%%%%%%%%%%%%%%%%%%%%%%%%%%%%%%%%%%%%%%%%%%

The limiting values of $p_c$ for small values of $\sigma$
can be understood as follows.
In the limit $\sigma \to 0$, the network tends to a set of
separated random regular (RR) graphs for each degree.
In this case there is no single giant component
but a set of giant components of each RR graph.
For the case when the minimum degree $m = 2$, the RR graph of the
smallest degree ($k = 2$) is composed of ``rings'' and therefore always critical,
which means the global connectivity of this component
of $k = 2$ collapse as soon as even a single node from this component is
removed. We can see this also from the fact that the critical node threshold
for the random regular graph of $k = 2$ is $p_c = 1/(k - 1) = 1$.
The giant component vanishes as soon as
a node in the last remaining component ($k = 2$)
containing the fraction $P(2)$ of nodes is removed.
Thus, $p_c = P(2)$ for $\sigma = 0$.
Above some positive finite value of $\sigma$, however,
the separated RR graphs become interconnected (hierarchically).
In this case, $\sigma > 0$,
the smallest degree component ($k = 2$) of the whole
network is connected to the $k = 3$ component.
This connection breaks up the giant component in the RR graph for $k = 2$,
since the $k = 2$ component is always critical.
The final giant component collapse thus comes when
the global connectivity of the $k = 3$ component is lost. The threshold for
the $k = 3$ component is $1/(k - 1) = 1/2$. Thus, the critical node threshold
at the collapse is expected to be $p_c = P(2) + P(3)/2$ for $\sigma > 0$.
(See Fig.\ \ref{fig:eigen_values} (a) and (b).)
Since for $\sigma = 0$, $p_c = P(2) $,
there is a discontinuity of $p_c$ for $m = 2$ at $\sigma = 0$.
Note that in Fig.\ \ref{fig:eigen_values} (a) and (b) for $m = 2$,
we only show the giant components for
$\sigma \gtrsim 0.35$ where the largest connected component spans the entire network.

For $m \ge 3$,
the last remaining giant component is
that of the smallest degree network
which is not critical for any non-zero value of $\sigma$.
The critical node threshold at the collapse of the giant component is,
therefore, $p_c = P(m)/(m - 1)$
(see Eq.~(\ref{eq:pc}), Figs.~\ref{fig:eigen_values} (c) and (d)).
We have also checked the continuity of the threshold as a function of $\sigma$ by numerically
calculating $p_c$ for $\sigma \gtrsim 0.01$.

It is clear that when we only consider the robustness against
targeted high degree node attack the limit $\sigma \to 0$
always gives the most robust structure,
which is the minimally interconnected RR graphs.

\subsection{Giant component collapse}
\label{ssec:GC collapse}

\begin{figure*}
 \includegraphics[clip,width=13cm]{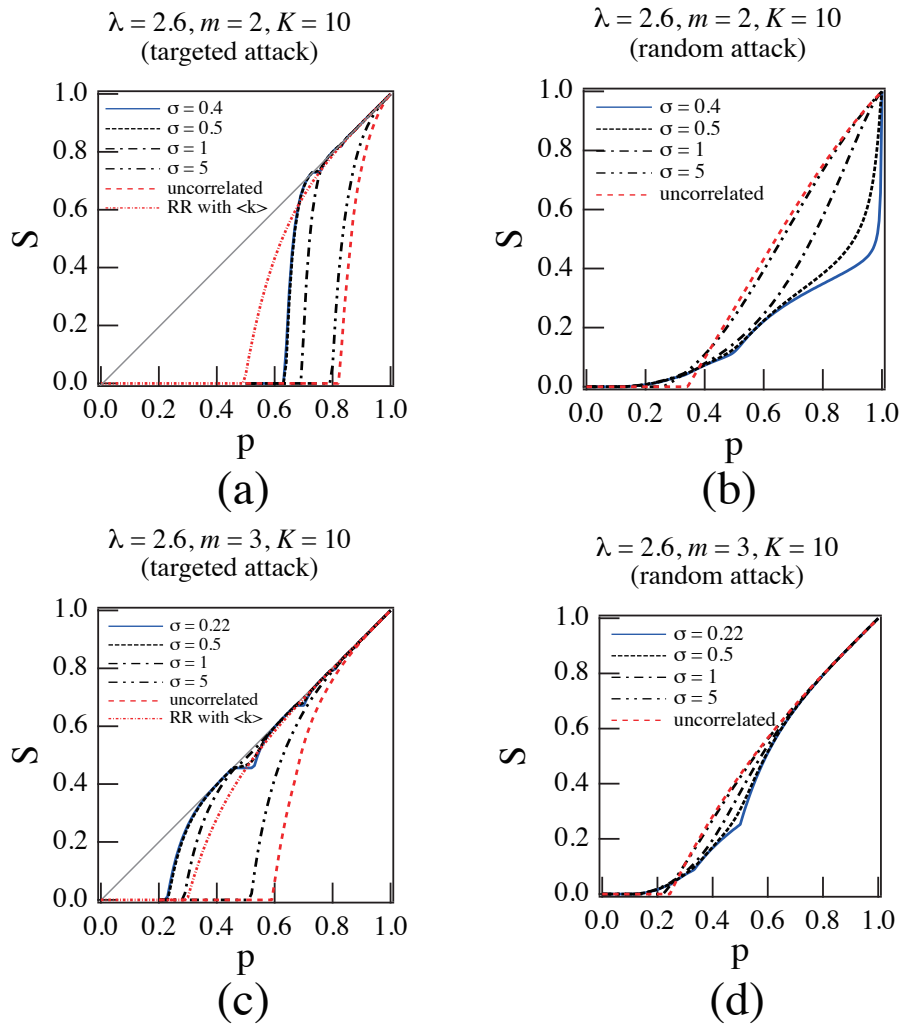}
 \caption{\label{fig:theor_GC_collapse}(Color online)
 Plots of the giant component fraction, $S$,
 as a function of remaining nodes, $p$, for several values of $\sigma$.
 The plots for the scale-free networks for $\lambda = 2.6$,
 $m = 2$, and $K = 10$ are (a) for targeted attack and (b) for random attack.
 The plots for the scale-free networks for $\lambda = 2.6$,
 $m = 3$, and $K = 10$ are (c) for targeted attack and (d) for random attack.
 For comparison, we also plot in (a) and (c)
 the curves of the giant component for uncorrelated networks
 and the curves for random regular (RR) networks
 where all degrees are the same as the average degree,
 $\langle k \rangle$, of the corresponding scale-free networks.
 The lines $S = p$ in (a) and (c) are guides for eyes
 and represent an optimal network.
 }
\end{figure*}

In Fig.\ \ref{fig:theor_GC_collapse},
we show the way of the giant component collapses
for various values of $\sigma$ 
for scale-free networks with an exponent, $\lambda = 2.6$.
Figures \ref{fig:theor_GC_collapse} (a) and (b) correspond
to a network with $m = 2$
and Figs.\ \ref{fig:theor_GC_collapse} (c) and (d) correspond
to a network with $m = 3$.
Note that for $m = 2$ the giant component corresponds to
a network composed of nodes mainly with the minimum degree, $m = 2$.
It is therefore at the edge of criticality
for extremely assortative correlation ($\sigma \to 0$).
This is the reason for the sudden collapse of the giant component
in the vicinity of infinitesimal removal of node by random attack,
which is represented by the solid curve for $\sigma = 0.4$
in Fig.\ \ref{fig:theor_GC_collapse} (b).

% %%%%%%%%%%%%%%%%%%%%%%%%%%%%%%%%%%%%%%%%%%%%%%
% Figure 5
% %%%%%%%%%%%%%%%%%%%%%%%%%%%%%%%%%%%%%%%%%%%%%%
% %%%%%%%%%%%%%%%%%%%%%%%%%%%%%%%%%%%%%%%%%%%%%%%%

From Figs.\ \ref{fig:theor_GC_collapse} (a) and (c) which represent the cases
of targeted attack, we can see that the strong assortative degree-degree
correlation that leads to the structure of weakly interconnected random
regular graphs yields much smaller values of $ p_c $,
which is $P(2) + P(3)/2$ for $m = 2$ and $P(3)/2$ for $m = 3$,
compared to the corresponding uncorrelated networks.
We also see from Figs.\ \ref{fig:theor_GC_collapse} (a) and (c)
that, for $p > p_c$, $S \approx p$ when $\sigma \lesssim 1$
until the sharp decrease in $S$ near $p_c$ sets in.
This means that in this case the removal of high degree nodes does not affect
the connectivity of the remaining giant component
and that the networks for $\sigma \lesssim 1$ have almost the maximum
robustness against targeted attack.

From Figs.\ \ref{fig:theor_GC_collapse} (b) and (d),
we can see that
the giant component decreases faster
in the early stages of the random node removal,
while the values of the critical node threshold are slightly
lower from those of random networks
due to the assortative degree-degree correlation.

The decrease of $\sigma$ has, therefore, opposite effects
in terms of the giant component collapse with respect to
targeted and random attacks.
For targeted attack, the collapse of the giant component
is maximally suppressed ($S \approx p$) for
small values of $\sigma \lesssim 1$,
while for random attack the collapse sets in earlier
compared to in the cases of uncorrelated (random) networks.
This fact means that
there must be an optimal value of $\sigma$ that considers both targeted and random attacks.
The structure of the network at this $\sigma $
suppresses the giant component collapse as much
as possible for targeted attack as well as
maintaining a sufficient fraction of giant component
for random attack.

\subsection{Robustness optimization}
\label{ssec:opt}

% %%%%%%%%%%%%%%%%%%%%%%%%%%%%%%%%%%%%%%%%%
% Figure 6
% %%%%%%%%%%%%%%%%%%%%%%%%%%%%%%%%%%%%%%%%%
\begin{figure}
 \includegraphics[clip,width=4.5cm]{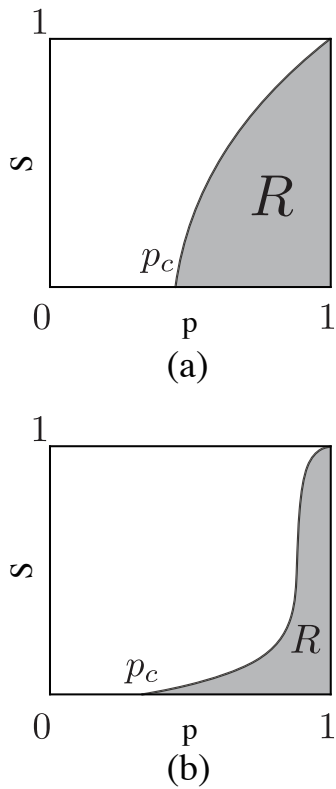}
 \caption{\label{fig:robustness_measure}(Color online)
 Schematic profiles of the giant component
 as a function of the remaining node fraction for the two typical cases.
 The case (a) has a larger value of the critical node threshold, $p_c$,
 than (b), but the giant component collapse for (a)
 occurs much slower than for the case (b).
 From the viewpoint of the global connectivity,
 the value of the area below $S(p)$ represented by $R$
 is a better measure of the robustness than the critical threshold, $p_c$.
 }
\end{figure}
% %%%%%%%%%%%%%%%%%%%%%%%%%%%%%%%%%%%%%%%%%%

To identify the optimal structure for both targeted and random attacks
we propose the following approach.
In Fig.\ \ref{fig:robustness_measure},
we show schematic profiles of the giant component fraction
as a function of $p$ for two possible scenarios motivated by the curves
appearing in Figs.~\ref{fig:theor_GC_collapse}(a) and (b).
The case represented by Fig.~\ref{fig:robustness_measure}(a)
has a larger value of the critical node threshold, $p_c$,
than that of Fig.~\ref{fig:robustness_measure}(b).
On the other hand a large fraction of the giant component collapse
after removal of a small fraction of nodes
for the case (b) in contrast to (a).
From the viewpoint of the macroscopic connectivity,
the value of the area under the curve $S(p)$ represented by $R$
is a better measure of the robustness,
as proposed by Schneider et al.\cite{Schneider:2011ip,Herrmann:2011hd},
compared to the critical node threshold, $p_c$,
and we apply in the following.
Note that this measure $R$ has the absolute upper bound of 0.5.

% %%%%%%%%%%%%%%%%%%%%%%%%%%%%%%%%%%%%%%%%%%
% Figure 7
% %%%%%%%%%%%%%%%%%%%%%%%%%%%%%%%%%%%%%%%%%%
\begin{figure*}
 \includegraphics[clip,width=12cm]{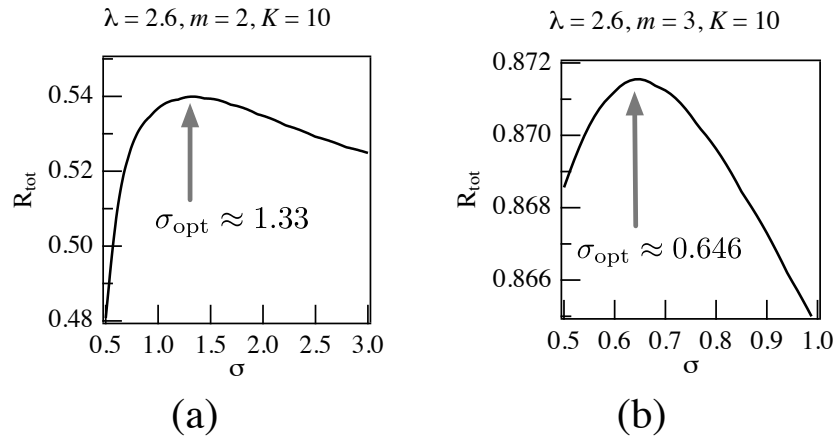}
 \caption{\label{fig:opt_sigma}
 The values of the total robustness measure $R_{\text{tot}}$
 as a function of $\sigma$ for two scale-free networks.
 Plot (a) is for a scale-free network
 with $\lambda = 2.6$, $m = 2$, and $K = 10$ and
 plot (b) is for a scale-free network
 with $\lambda = 2.6$, $m = 3$, and $K = 10$.
 The numerically obtained optimal values of $\sigma$
 that maximize $R_{\text{tot}}$ are $\sigma = 1.33$ for $m = 2$
 and $\sigma = 0.646$ for $m = 3$.
 }
\end{figure*}
% %%%%%%%%%%%%%%%%%%%%%%%%%%%%%%%%%%%%%%%%%%%

Since we are considering the total robustness
against both targeted and random attacks,
we define the total robustness measure $R_{\text{tot}}$ as the sum of
the robustness measure for targeted attack and that for random attack.
Defining the total measure as the sum of the measures against
both attacks is also found in earlier literature
\cite{Paul:2004br,Tanizawa:2005fd,Tanizawa:2006dn}.
Note that the maximum value of $R_{\text{tot}}$ equals to unity for
networks with complete robustness in which $p_c = 0$ and $S = p$ for
both targeted and random attacks.

Figure \ref{fig:opt_sigma} shows $R_{\text{tot}}$ as a function of
$\sigma$ for two scale-free networks.
The plot in Fig.~\ref{fig:opt_sigma}(a) is for a scale-free network
with $\lambda = 2.6$, $m = 2$, and $K = 10$.
The measure, $R_{\text{tot}}$,
reaches the maximum value of approximately $0.54$ at
$\sigma_{\text{opt}} \approx 1.33$.
Figure \ref{fig:opt_sigma}(b) is for a scale-free network
with $\lambda = 2.6$, $m = 3$, and $K = 10$.
The measure, $R_{\text{tot}}$, reaches
the maximum value of approximately $0.8715$ at
$\sigma_{\text{opt}} \approx 0.646$.
Noticing that the limit $\sigma \to 0$ leads to the separated
random regular (RR) networks and that $\sigma$ is
the measure of the maximum degree difference of connected nodes
(See Eq.~(\ref{eq: trans joint deg mat})),
the fact that $\sigma_{\text{opt}} \approx 1$ for both cases
of $m = 2$ and $m = 3$ indicates that the optimal network
structure is the one where
most of the $k$-degree nodes are connected with each other
and only a small fraction of remaining $k$-degree nodes are connected
with $(k-1)$ or $(k+1)$-degree nodes.
We find here again the onion-like structure.
The reason that $\sigma_{\text{opt}}$ for $m = 2$ is
slightly larger than the one for $m = 3$ is due to
the criticality of the smallest degree component of $k = 2$.

% %%%%%%%%%%%%%%%%%%%%%%%%%%%%%%%%%%%%%%%%%%%
% Figure 8
% %%%%%%%%%%%%%%%%%%%%%%%%%%%%%%%%%%%%%%%%%%%
\begin{figure*}
 \includegraphics[clip,width=12.5cm]{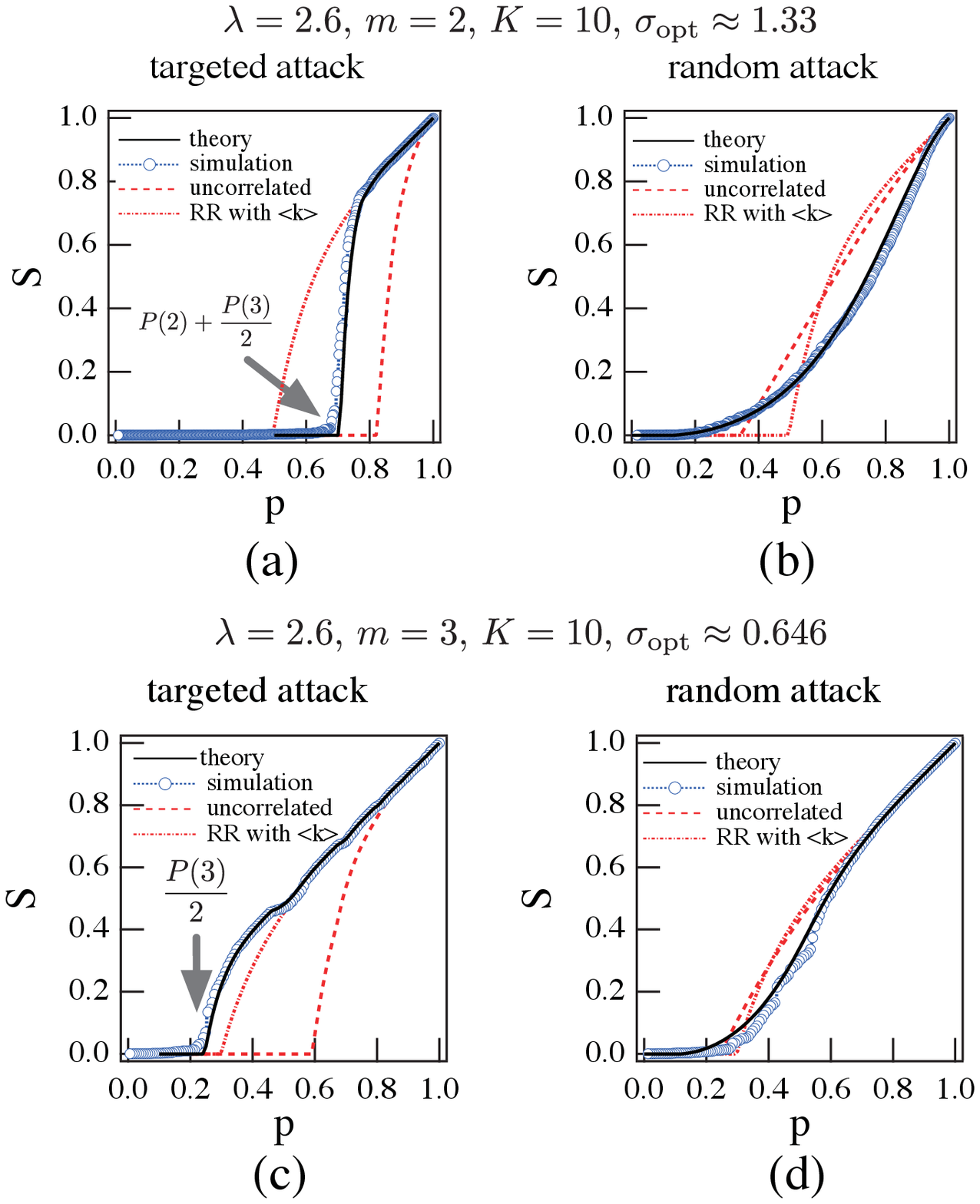}
 \caption{\label{fig:opt_sigma_GC}(Color online)
 Plots of the optimal giant component, $S$,
 as a function of $p$ for scale-free networks
 with $\lambda = 2.6$ and $K = 10$.
 For plots (a) targeted attack and (b) random attack,
 $m = 2$ and the optimal value is $\sigma = 1.33$. For
 plots (c) targeted attack and (d) random attack, $m =3$ and
 the optimal value is $\sigma = 0.646$.
 In all plots, the theoretical values for the giant component
 are represented by full curves.
 The critical node thresholds for targeted attack
 are $P(2) + P(3)/2$ for $m = 2$ and $P(3)/2$ for $m = 3$.
 We also plot, for comparison, the curves
 for the corresponding uncorrelated scale-free network with
 the same values of parameters (dashed curves)
 and for the RR network with the same degree
 as the average degree of the corresponding scale-free network
 (dotted curves).
 The (blue) circles are obtained
 from simulation of a single realization
 for each of the optimal networks generated from the joint degree-degree
 matrix, Eq.\ (\ref{eq: trans joint deg mat}),
 for the optimal value of $\sigma = 1.33$
 with $N = 6993$ and $m = 2$ and for the optimal value of $\sigma = 0.646$
 with $N = 2795$ and $m = 3$.
 }
\end{figure*}
% %%%%%%%%%%%%%%%%%%%%%%%%%%%%%%%%%%%%%%%%%%%

In Fig.\ \ref{fig:opt_sigma_GC}, we plot
the giant components for targeted and random attacks
as a function of $p$ for the optimally correlated scale-free networks
with $\lambda = 2.6$ and $K = 10$
for the cases $m = 2$ ($\sigma = 1.33$) and $m = 3$ ($\sigma = 0.646$).
In all plots, the theoretical values of the giant component fraction
are represented by full curves.
The critical node thresholds for targeted attack are
$P(2) + P(3)/2$ for $m = 2$ and $P(3)/2$ for $m = 3$, respectively.
For comparison, we also plot the curves for
the corresponding uncorrelated scale-free network with
the same parameters and for the RR network of the same degree
as the average degree of the corresponding scale-free networks.
These results show that the robustness of scale-free networks
against targeted attack can be significantly improved
up to nearly maximal by taking the structure of weakly
interconnected RR graphs (onion-like structures)
without much undermining their intrinsic robustness
against random failure.

For testing our theoretical considerations,
we also simulate actual networks according to the joint degree matrix,
Eq.\ (\ref{eq: trans joint deg mat}), with the optimal values of $\sigma$,
which are $1.33$ for $m = 2$ and $0.646$ for $m = 3$.
The circles in Fig.\ \ref{fig:opt_sigma_GC} are obtained
by direct node removal from the simulated optimal networks.
For each realization, the number of nodes
for $m = 2$ is 6993 and for $m = 3$ is 2795.
The agreement between the simulation results and the
theoretical calculations is excellent.

\section{Summary}
\label{sec:summary}

As a strong candidate for the optimal structure
against both types of attacks, random and targeted,
with a given degree distribution,
the structure consisting of hierarchically interconnected
random regular graphs is proposed and
thoroughly investigated based on exact analytical expressions.
This network structure has a close relationship with the ``onion-like structure''
found by Schneider et al.~\cite{Schneider:2011ip,Herrmann:2011hd} using
numerical simulations and
exhibits an extremely assortative degree-degree correlation,
in which a node of certain degree has
a strong tendency to be linked with
nodes of the same degree.
We derive a set of exact expressions that enable us to calculate
the critical node threshold
and the giant component fraction for arbitrary types of node removal,
in which the degree-degree correlation is fully incorporated.
To test the robustness of this structure,
we apply the theory to the case of scale-free networks that have
a well-known vulnerability against targeted attack.
The results show that the vulnerability of a scale-free network
can be significantly improved by taking the network structure
proposed here without much undermining its almost complete robustness
against random attack.
We also investigate the detail of the robustness enhancement
of scale-free networks due to assortative degree-degree correlation
by introducing a joint degree-degree probability matrix
that interpolates between an uncorrelated
network structure and the structure with strong assortativity
by tuning a single control parameter.
The optimal values of the control parameter that maximize the robustness
against simultaneous random and targeted attacks are also determined
and those optimal values support the maximal robustness of the ``onion-like structure.''
Our analytical calculations are supported by numerical simulations.

\section*{Acknowledgment}

This work is supported by the Grant-In-Aid for Scientific Research (C)
from the Japan Society for the Promotion of Science, No.\ 20540382.
We also wish to thank the ONR, DTRA, DFG, EU project Epiwork, the LINC EU project,
and the Israel Science Foundation for financial support.

%%%%%%%%%%%%%%%%%%%%%%%%%%%%%%%%%%%%%%%%%%%%%%%%%%%%%%%%%%%%%%%%%%%%%%%
% References
%%%%%%%%%%%%%%%%%%%%%%%%%%%%%%%%%%%%%%%%%%%%%%%%%%%%%%%%%%%%%%%%%%%%%%%
% \bibliography{opt_corr_bib} % Produces the bibliography via BibTeX.

%merlin.mbs 2010-03-15 4.21a (PWD, AO, DPC)
%Control: key (0)
%Control: author (8) initials jnrlst
%Control: editor formatted (1) identically to author
%Control: production of article title (-1) disabled
%Control: page (0) single
%Control: year (1) truncated
%Control: production of eprint (0) enabled
%

\end{document}